\documentclass[final,authoryear,1p,times]{elsarticle}
\usepackage{amssymb}
\begin{document}
\begin{frontmatter}
\title{Neutrinos from photo-hadronic interactions in Pks2155-304}
\author[bochum,gbg]{Julia K.\ Becker}
\author[erlangen]{Athina Meli}
\author[bonn1,bonn2,ala1,ala2,karls]{Peter L.\ Biermann}
\address[bochum]{Ruhr-Universit\"at Bochum, Fakult\"at f\"ur Physik \& Astronomie, Theoretische Physik IV,
 44780 Bochum, Germany}
\address[gbg]{G\"oteborgs Universitet, Institution f\"or Fysik, 41296
  G\"oteborg, Sweden}
\address[erlangen]{Erlangen Center for  Astroparticle Physics, Friedrich-Alexander
 Univ. Erlangen-N{\"u}rnberg, Germany}
\address[bonn1]{MPI for Radioastronomy, Bonn, Germany}
\address[bonn2]{Bonn University, Germany}
\address[ala1]{Dept. of Phys. \& Astr., Univ. of Alabama, Tuscaloosa, AL, USA}
\address[ala2]{Dept. of Phys., Univ. of Alabama at Huntsville, AL, USA}
\address[karls]{FZ Karlsruhe, and Phys. Dept., Univ. Karlsruhe, Germany}

\begin{abstract}
\noindent
The high-peaked BL Lac object \mbox{Pks2155-304} shows high variability at
multiwavelengths, i.e.\ from optical up to TeV energies. A giant flare
of around 1 hour at X-ray and TeV energies was observed in 2006
\cite{aharonian_pks2155_2007}. In this context, it is essential to
understand the physical processes in terms of the primary spectrum and
the radiation emitted, since high-energy emission can arise in both
leptonic and hadronic processes. 
In this contribution, we investigate the possibility of neutrino
production in photo-hadronic interactions. 
In particular, we predict a direct correlation between optical and TeV
energies at sufficiently high optical radiation fields. We show that
in the blazar Pks2155-304, the optical emission in the low-state is
sufficient to lead to photo-hadronic interactions and therefore to the
production of high-energy photons. 
\end{abstract}
\begin{keyword}
neutrinos \sep Pks2155-304 \sep photo-hadronic interactions
\end{keyword}

\end{frontmatter}

\noindent
\parindent=0.2cm
\section{Very high-energy emission from blazars}
The very first detection of a blazar at TeV photon energies happened
in 1992, when Mkn421 was observed with the HEGRA telescopes
\citep{punch1992}. Since then, large progress has been
made detecting more and more active galactic nuclei (AGN) with detailed spectral
features. As of today, 26 blazars as well as 2 radio galaxies of
Fanaroff-Riley type\footnote{The radio galaxies are M 87
and Cen A, see \cite{roberts_homepage} for a complete list of known
sources} are known to emit at very
high-energy (VHE, i.e.~$E_{\gamma}>100$~GeV). Blazars typically show
a spectrum with a
double-hump structure, i.e.\ a low-energy peak at radio to optical
wavelengths (or X-ray) and a high-energy peak at X-ray (or gamma-ray) to TeV energies. The
entire emission is believed to arise in the
relativistic shock fronts in AGN jets, where charged particles are
accelerated to ultra high-energies (UHE) \citep[e.g.]{mbq2008,biermann_review}. Then, the low-energy component is
 due to synchrotron radiation by the accelerated electrons.
Several processes can produce VHE
radiation: With relativistic electrons accelerated in the shock fronts
of the blazar's jet, synchrotron photons are produced. The latter are
scattered to VHE via the Inverse Compton process with the primary
electrons. Hadronic Cosmic Rays (CRs) are believed to be accelerated
along with the electrons, up to $10^{21}$~eV. These CRs can radiate at VHE through
synchrotron radiation, or interactions with ambient matter or photon
fields. Here, we consider the possibility of CR interactions with
photon fields.

Blazars are observed to be highly variable at all
wavelengths. The source of this variability is still
under discussion, ranging from instabilities in the accretion disk,
see e.g.\ \cite{wiita2006}, to precession of the jet, see
e.g.\ \cite{gergely_biermann2009}. Other explanations are jets with
density inhomogeneities or turbulent jets, see
\cite{romero1995} and references therein. Such phenomena lead to the
variation of the flux at low energies (between radio and X-ray). The
variability at the higher energies (X-ray to TeV energies) depends on
the VHE radiation origin: 
\begin{enumerate}
\item {\it Inverse Compton scattering} can arise in two different scenarios:
On the one hand, it can be due to the interaction between the accelerated electrons
with the synchrotron photon field (``Synchrotron Self Compton'', SSC).
This leads to a direct correlation between the low- and high-energy
humps of the spectrum. On the other hand, if external
photons are boosted to VHE by electrons, such a correlation is not
necessary. The high-energy hump would then show a variability
uncorrelated to the low-energy spectrum.
\item {\it Hadronic interactions} can lead to the production of VHE
  photons through the production of $\pi^{0}$, as neutral pions decay into
  two VHE photons. Here, a correlation between the low- and
  high-energy hump is not necessary. It can, however, be present if
  the hadronic CRs interact with the synchrotron photon field
  itself. In the case of proton-proton interactions, no correlation
  would be present. In addition, synchrotron proton radiation can
  contribute at VHE.
\item {\it A combination of hadronic and leptonic VHE emission} 
is likely to be present in many cases, as protons and electrons are
co-accelerated in the jet's shock fronts. The continuous interaction of the different radiation
fields and charged particle populations may lead to a series of loops
of VHE production. This may lead to a strongly non-linear correlation
between the intensity of different wavelengths. A description of
repeated leptonic loops is given in \cite{kellermann1969}. The same
mechanism may work for hadronic loops or a combination of both, always
assuming a low-energy target photon field.
\end{enumerate}

As the synchrotron field from primary electrons is a good target for
Inverse Compton scattering, a direct correlation between low- and
high-energy photons is usually interpreted as such. In so-called
orphan flares, a strong variability at VHE is observed in the
absence of low-energy variability. This in turn is often taken to be
an indication as a result of hadronic interactions, as these do not
require an intensity change at low energies. In this paper, we present
a model where variability at low-energies leads to a correlated signal
at high-energies. 


\section{Photo-hadronic interactions in blazars}
When CRs are accelerated to UHE, they can interact with the present
photon fields via the $\Delta-$resonance:
\begin{equation}
p\,\gamma \rightarrow \Delta^{+} \rightarrow \left\{\begin{array}{ll}p\,\pi^{0}\\n\,\pi^{+}\end{array}\right.
\end{equation}
The decay of the $\Delta-$resonance leads to the production of charged
and neutral pions, which decay into high-energy neutrinos and
photons, respectively \citep[e.g.]{julia_review}):
\begin{eqnarray}
\pi^{+}&\rightarrow& \mu^{+}\,\nu_{\mu}\rightarrow
e^{+}\,\nu_{e}\,\overline{\nu}_{\mu}\,\nu_{\mu}\; \\
\pi^{0}&\rightarrow& \gamma\,\gamma\,.
\end{eqnarray}
The probability for $\pi^{+}$ production is  $2/3$,
while $\pi^{0}$ are produced in $1/3$ of the cases. Around $1/5$ of
the proton energy goes into the pion, and the decay products carry
equal energy, so that a fraction of 0.25 (0.5) is transferred to each
neutrino (photon). Furthermore, we need to consider that the neutrinos
oscillate from 2 muon neutrinos and one electron neutrino to one
electron, one muon and
one tau neutrino.
Considering these factors, the high-energy
photon and neutrino spectra are connected as\footnote{for a detailed
  description, see \cite{francis_review2008}.}
\begin{equation}
\frac{dN_{\nu}}{dE_{\nu}}\approx \frac{1}{8}\cdot
\frac{dN_{\gamma}}{dE_{\gamma}}\,.
\label{VHE_nus_gammas:equ}
\end{equation}

Photo-hadronic interactions occur under the following conditions:
\begin{enumerate}
\item  Protons need to be accelerated to sufficiently high energies.
\item The photon field must be dense enough for the protons to
  interact, i.e.\ we need an optical depth for proton-photon
  interactions of $\tau_{p\gamma}\sim1$ or larger.
\end{enumerate}
The first condition is believed to be fulfilled, as AGN jets enable
proton acceleration up to $10^{21}$~eV, see
e.g.\ \cite{mbq2008,biermann_review}. An approximation of the optical depth
for photo-hadronic interactions in knots of blazars is given by
\cite{becker_biermann2009},
\begin{equation}
\tau_{p\,\gamma}\approx 0.6\cdot \left(\frac{L_{\rm opt}}{3\cdot 10^{45}\,{\rm
      erg/s}}\right)
\cdot \left(\frac{10}{\Gamma}\right)\cdot \left(\frac{\theta}{0.1}\right)\cdot \left(\frac{\epsilon_{\rm knot}}{10} \right) \cdot \left(\frac{z_{j}}{3000\,r_g}\right)^{-1}\cdot
  \left(\frac{\nu}{4.6\cdot 10^{5}\,{\rm GHz}}\right)^{-1}\,.\label{opt_depth_rband:equ}
\end{equation}
Here, $L_{\gamma}$ is the luminosity observed from the blazar at
a given frequency $\nu$. Further, $\Gamma\approx 10$ is the Lorentz factor
of the shock and $\theta\approx 0.1$ is the
jet's opening angle. The parameter $\epsilon_{\rm knot}$
gives the fraction of luminosity coming from a single knot in the
jet. The value $\epsilon_{\rm knot}=0.1$ indicates that the luminosity
comes from 10 knots, which is the order of magnitude to be
expected. Finally, $z_{j}$ is the distance of the knot along the
jet. The first shock is expected to be present at around $3000\,$
gravitational radii, $r_{g}$, from the foot of the jet \citep{marscher2008}. As the optical
depth drops with the distance from the center and we need those strong
shocks to have efficient neutrino production, this first strong shock
provides the best environment for the production of secondaries. The
flaring time scale of Pks2155-304 also suggests that the emission
happens in a compact region, underlining the fact that it could be due
to photohadronic interactions.


\section{Multiwavelength observations of Pks2155-304}
Pks2155-304 is a blazar at a redshift of z = 0.116, located in the southern 
hemisphere, at \mbox{(RA, Dec) = (21h58'52.0'', -30$^{\circ}$13'32'')} in J2000
coordinates \citep{fey2004}. 
It was the 5th source reported at $\gamma$-rays \citep{chadwick1999}, with a steady-state 
emission of around 20\% of the Crab nebula. Pks2155-304 shows a
very rapid variability, with extreme flares, the most intense one
observed in 2006 \citep{aharonian_pks2155_2007}, with an increased
intensity of more than one order of magnitude at TeV energies. This
extraordinary flare lasted about $60$~minutes and showed a peak intensity of
$7$ times the Crab nebula. CANGAROO-III was triggered by the
H.E.S.S.\ observations and detected the flaring state at complementary
times \cite{sakamoto2008}. During this flaring state, no other wavelength
  bands were observed.

In a later flare, a multiwavelength campaign of Chandra and
H.E.S.S.\ was arranged to study the correlations between the different
wavelength bands  \citep{costamante2008}. Here, X-ray and TeV variabilities were
more moderate, i.e.\ within a factor of 3, and optical emission varied
within a factor of 2. For this flare, it could be shown that TeV and X-ray
luminosities seem to be connected as \citep{costamante2008}
\begin{equation}
L_{\rm TeV} \propto L_{\rm X-ray}^{3}\,.
\end{equation}

The low-state of Pks2155-304 was monitored in a
multiwavelength campaign in 2008 \cite{aharonian_pks2155_2009}. Four
wavelength bands were observed simultaneously, namely at optical (ATOM), X-ray\\ (RXTE \& Swift),
$100$~MeV to $100$~GeV (Fermi-LAT) and \\ \mbox{$>200$~GeV} (H.E.S.S.) energies. While a strong correlation between
optical and VHE emission was found, X-ray energies do not correlate with
the highest energies. This behavior in the ground state is fundamentally different from the observed correlation between X-ray and TeV in the flaring state. The spectral index behavior observed by Fermi is
around $E^{-2\pm0.5}$, where the deviation $\pm 0.5$ corresponds to
the temporal change in the spectral index with time. 

\section{Neutrinos from Pks2155-304}
The neutrino spectrum from proton-photon interactions in blazars 
basically follows the primary proton spectrum above a break energy
$E_{\nu}^{\rm break}$, where the latter is given by the kinematic condition for
producing a $\Delta-$resonance \citep[e.g.]{julia_review}:
\begin{equation}
E_{\nu}^{\rm break}=\frac{\Gamma^2}{(1+z)^{2}}\cdot
\frac{m_{\Delta}^{2}-m_{p}^{2}}{80\cdot E_{\gamma}}
\label{ebreak1:equ}
\end{equation}
Here, both the break energy $E_{\nu}^{\rm break}$ and the photon
energy of the target photons, $E_{\gamma}$, are given in the observer's frame at Earth. In the
equation, it was considered that the
neutrino carries $1/20$th of the proton's energy, $E_{\nu}=E_p/20$. We
can now express Equ.\ (\ref{ebreak1:equ}) numerically as 
\begin{equation}
E_{\nu}^{\rm break}=624\,\mbox{GeV}\,\cdot \left(\frac{\Gamma}{10}\right)^{2}
\cdot
\left(\frac{E_{\gamma}}{MeV}\right)^{-1}\cdot 
\left(\frac{1+z}{1.116}\right)^{-2}\,.
\end{equation}
Below that break energy, the neutrino spectrum is about one power
flatter than the proton spectrum. 

Given the observation of the low-state of Pks2155-304 in 2008
\cite{aharonian_pks2155_2009}, we can calculate the optical depth for
photo-hadronic interactions. Here, we investigate if the observed spectral
energy distribution (SED) of photons is a good target field. Using that the frequency of the R-band
photons is $\nu= 4.6\cdot 10^5$~GHz, 
we can calculate the luminosity of the source at a given frequency
$\nu$ from the observed flux,
$F_{\gamma}$, and from the distance of
Pks2155-304, $d_L$, is given by the blazar's redshift\footnote{we use a $\Lambda$CDM
  cosmology with $(\Omega_{m},\,\Omega_{\Lambda},\,h) =
  (0.3,\,0.7,\,0.7)$},
\begin{equation}
L_{\rm opt}=3\cdot 10^{45}\cdot \left(\frac{F_{\gamma}}{10^{-10}\,{\rm erg/s/cm}^2}\right)\,{\rm erg/s}\,.
\end{equation}
Here, the reference value corresponds to the flux in the R-band, $\nu= 4.6\cdot 10^5$~GHz.
Given Equ.\ (\ref{opt_depth_rband:equ}), we can derive the optical
depth for proton-photon interactions during the low-state observation
of Pks2155-304 for the multiwavelength data.

\begin{figure}[!t]
\centering
\includegraphics[width=\linewidth]{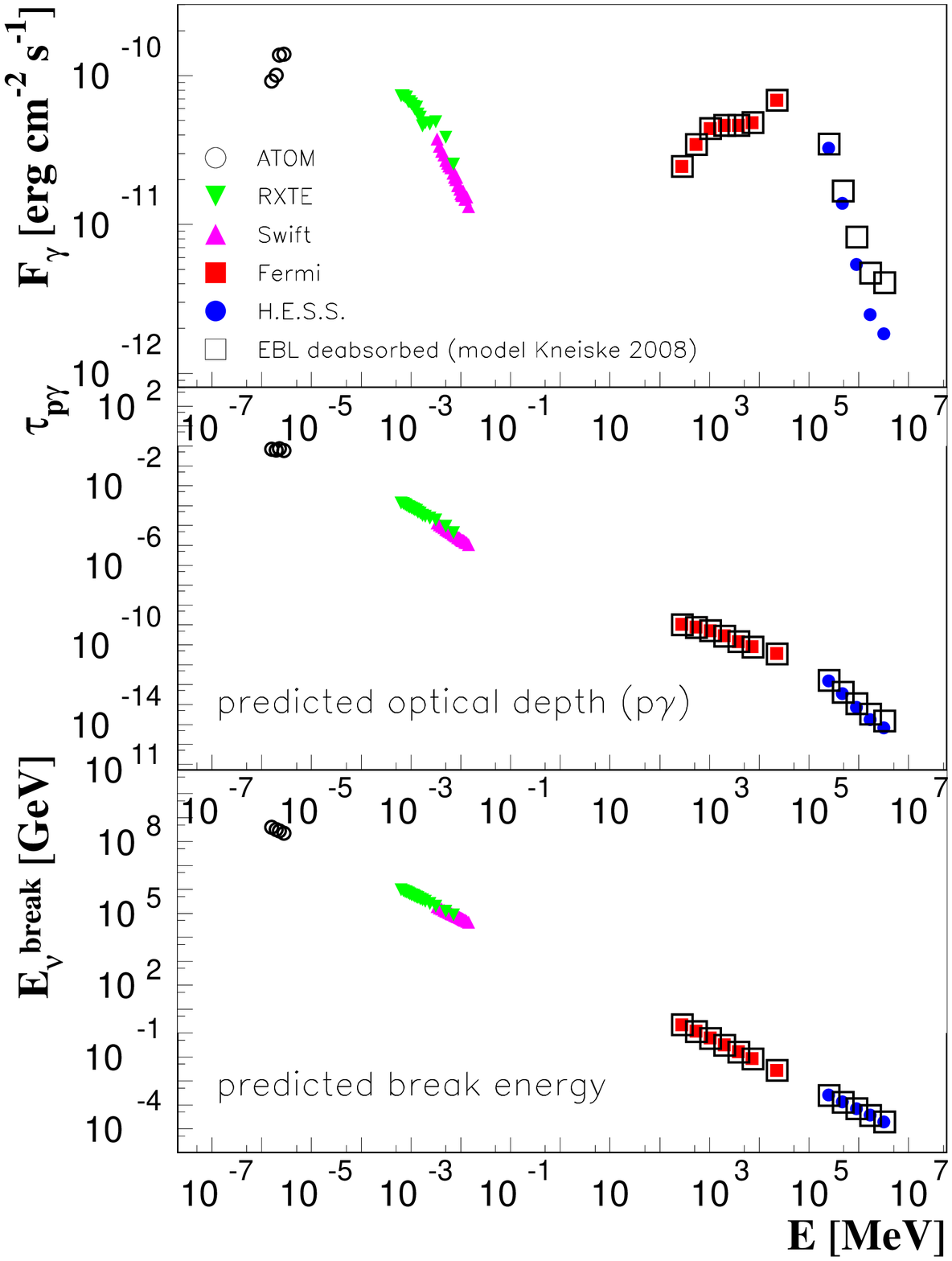}
\caption{(a) Spectral energy distribution of photons from Pks2155-304
  in the low state - data from \citep{aharonian_pks2155_2009}. Open squares represent the flux at TeV energies without absorption due to the extragalactic background light, using the model by \cite{tanja_lower_limit_ebl} for deconvolution. (b) Photohadronic
  optical depth using the SED as the interacting photon field. (c)
  Break energy of the neutrino spectrum for proton interactions with
  photons at a given energy.}
\label{sed:fig}
\end{figure}

Figure \ref{sed:fig} shows three panels: the upper one shows the SED
as measured in the low-state \citep{aharonian_pks2155_2009}. We can
now investigate how efficient proton-photon interactions at different
wavelengths can be and what the neutrino spectrum is. The second panel
of Fig.\ \ref{sed:fig} shows the optical depth for proton-photon
interactions. At optical wavelengths, the optical depth is close to
unity (around $\tau_{p\gamma}\sim 0.7$), and it decreases significantly
towards higher energies. Thus, we conclude that proton-photon
interactions are efficient with optical photons, but not with
higher energy photons. This implies that the X-ray luminosity of the
source does not need to be correlated with the optical and high-energy
emission, just as it is observed in the low-state. The lowest panel of Fig.\ \ref{sed:fig}
displays the break energy of the neutrino spectrum for photohadronic
interactions for the given wavelength. It is shown that at optical
wavelengths, the break energy lies around $\sim 10^{8}$~GeV and this
value decreases towards higher energy target photon fields. This is a
general result, since the neutrino break energy is inversely
proportional to the target photon field energy.

\begin{figure}[!t]
\centering
\includegraphics[width=\linewidth]{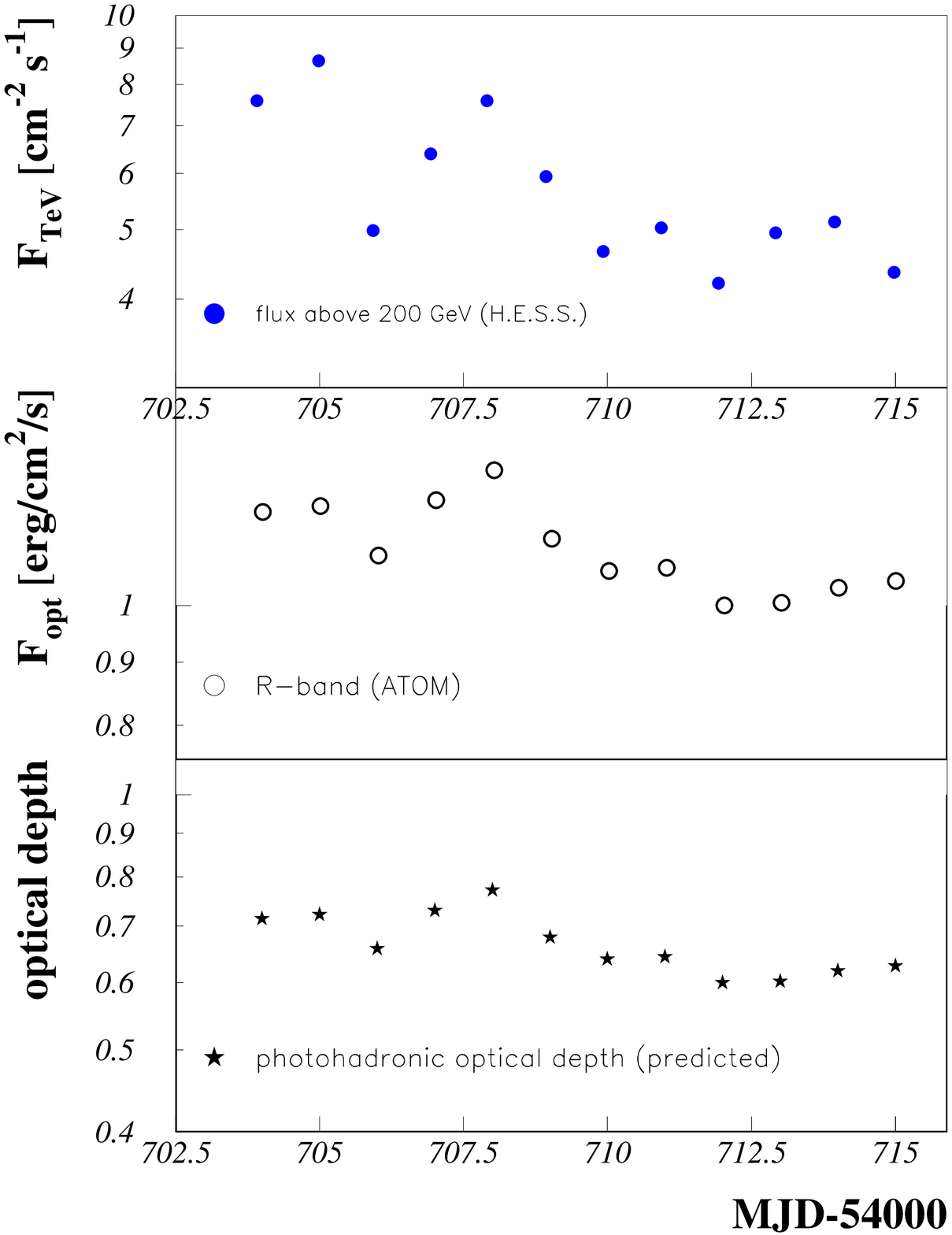}
\caption{(a) TeV lightcurve observed by
  H.E.S.S. \citep{aharonian_pks2155_2009}. (b) R-band lightcurve
  observed by ATOM \citep{aharonian_pks2155_2009}. Predicted variation of the photohadronic optical depth with time.}
\label{tau_time:fig}
\end{figure}

The temporal dependence of the photo-hadronic optical depth with
R-band photons, derived from the low-state measurements of Pks2155-304
\citep{aharonian_pks2155_2009}, is shown in Fig.\ \ref{tau_time:fig}. The upper panel
shows the observed TeV emission, which correlates to the emission in
the R-band (middle panel). The photo-hadronic optical depth is displayed in the
lower panel. The optical depth varies in coincidence with the R-band
signal, between $0.6<\tau_{p\gamma_{\rm opt}}<0.8$, yielding a reasonable
efficiency for the production of VHE photons and neutrinos.

Given an average flux in the R-band of $\sim 1.1\cdot 10^{-10}$~erg/cm$^2$/s, the mean optical
depth is $\sigma_{p\gamma_{\rm opt}}\sim 0.7$. 
Considering B- and V-bands would increase the optical depth
further. With these \mbox{values} close to unity, the
observed flux of VHE photons can be due to proton interactions with optical
photons. The observed correlation between R-band and TeV
emission can therefore be interpreted as a result of photo-hadronic
interactions. 

Given an optical depth of around unity, and assuming VHE photons
to come from photo-hadronic interactions, we can calculate the
resulting neutrino flux for the low-state of Pks2155-304 using
Equ.\ (\ref{VHE_nus_gammas:equ}). Here, we use the Fermi measurements
as a reference flux. The reason for not using H.E.S.S.\ measurements
is that here, we can expect to see effects like absorption, while the
spectrum as observed by Fermi should be unaffected. The spectral
behavior as observed by Fermi, close to $dN_{\gamma}/dE_{\gamma}\propto
E^{-2}$, seems to mirror the primary spectrum and can therefore be
interpreted as unaffected \cite{mbq2008,bednarz_ostrowski1998}. The resulting neutrino flux is given as the
black, solid line in
Fig.\ \ref{nuflux:fig}. Also shown is the predicted 1-year sensitivity
for KM3NeT, a cubic kilometer sized neutrino detector planned to be
instrumented in the Mediterranean Sea. 
While the energy output in neutrinos is in general high enough to be
observed by large volume neutrino telescopes, the challenge will be to
have good sensitivities at energies as high as $10^{8}$~GeV. 
In \citep{ic40_icrc}, it is
reported that also IceCube is sensitive to sources in the southern
hemisphere, if only considering the highest energy events and thereby
rejecting the large background of atmospheric muons. IceCube's
sensitivity of the 40-string configuration is around
\begin{equation}
\frac{dN_{\nu}}{dE_{\nu}}=5\cdot
10^{-8}\,\mbox{GeV/s/cm}^2\,\left(\frac{E}{GeV}\right)^{-2}
\label{ic40_sens_pks2155:equ}
\end{equation} 
in the energy range of $\sim 10^{6}$~GeV~$-10^{9}$~GeV. The
sensitivity at these high energies is very well-suited for the
observation of Pks2155-304. In order to observe the low-level state,
the sensitivity needs to be increased by a factor of 10, which seems
possible considering that Equ.\ (\ref{ic40_sens_pks2155:equ}) refers to 330 days of the
IceCube 40 string configuration. The final IceCube detector with 86 strings, expected to run
around 10 years or more, is expected to yield significantly improved
sensitivities. This implies that such a flux from Pks2155-304 should
be detectable within the first years of IceCube.
\subsection{Flux variability in high-energy photons and neutrinos}
\begin{figure}[!t]
\centering
\includegraphics[width=\linewidth]{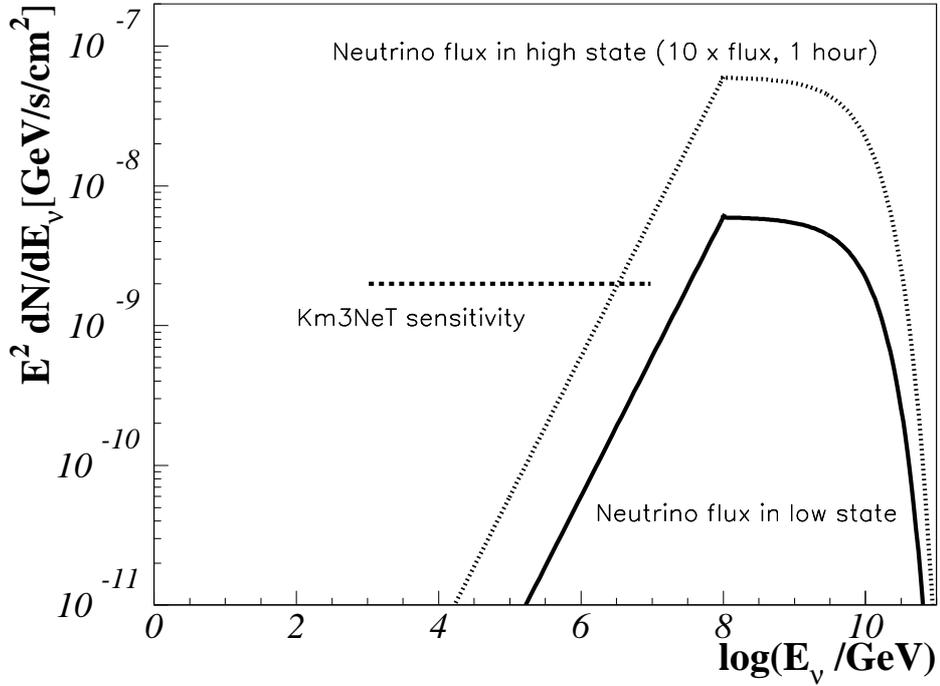}
\caption{Expected neutrino flux from photo-hadronic interactions in Pks2155-304. The solid line is the
  predicted average neutrino flux in the low-state as observed in
  2008, using an $E^{-2}$ primary spectrum. The dotted line shows the
  expected emission level in the high-state of
  Pks2155-304 during the flare observed in \cite{aharonian_pks2155_2007}.}
\label{nuflux:fig}
\end{figure}

In the previous section, we investigated the possible neutrino flux
from Pks2155-304 in the low-state. This component should be present
permanently. The extreme variability of this source at the highest
energies indicates that, in addition to this permanent component, one
can expect an enhanced contribution during the time of flaring
states. We investigate this fact at the example of the giant flare of Pks2155-304
observed by the H.E.S.S.\ experiment in 2006
\cite{aharonian_pks2155_2007}. Here, the VHE flux of Pks2155-304
showed increased activity by a factor of $\sim 10$ for about one hour. If we assume
that the VHE emission is due to photo-hadronic interactions, we can
calculate the corresponding neutrino flux according to
Equ.\ \ref{VHE_nus_gammas:equ}. The result is shown as the 
dotted lines in
Fig.\ \ref{nuflux:fig}. The neutrino emission, just as the VHE photon component, is
higher by approximately a factor of ten compared to the low-state. 

Even in this flaring state, the increased intensity can arise from an
enhanced optical emission. On the other hand, X-ray luminosities
are observed to be coupled to the VHE luminosity with a correlation of
$L_{VHE}\propto L_{X-ray}^{3}$ \cite{costamante2008}. The question is
how to explain this very strong dependence of the VHE emission with
the X-ray signal. One approach is the production of VHE radiation in
several loops of interactions as discussed in \cite{kellermann1969}
for Inverse Compton scattering. The VHE
luminosity is assumed to be produced by a series of $n$ loops of interactions between
the primary electrons and the secondary synchrotron photons,
\begin{equation}
L_{VHE}=L_{synch}\cdot \sum_{i=1}^{n} \xi^{i}\,,
\end{equation}
Here, $L_{synch}$ is the synchrotron radiation field and
$\xi=L_{VHE,0}/L_{synch}$. If the initially produced VHE radiation $L_{VHE,0}$ is
larger than the synchrotron field, the series diverges for infinite
loops $n\rightarrow \infty$. However, if the VHE radiation is cutoff at some point when
going into the Klein-Nishina limit, we have a maximum number $n$ which
can be small. If this happens at $n=3$, we can reproduce the cubic
correlation observed between VHE and
synchrotron emission for the case of $n=3$ and $\xi>1$. A similar effect
could be present if the VHE radiation was not due to Inverse Compton
scattering but due to $\pi^{0}$ decays. A detailed study of hadronic
loops and combinations of leptonic and hadronic loops is in
preparation. 


\subsection{Flaring and permanent neutrino emission states}
As discussed above, an enhanced flux of photons can
arise from photo-hadronic interactions, leading to the coincident
production and emission of neutrinos. In the giant flare, the flux
increase was about an order of magnitude for the duration of $\sim 60$~min.
Note, however, that this enhanced flux is only present in the one hour when the
source is flaring at very high photon energies. We can perform a simple exercise at what
point flares can be more significant in neutrino detectors as KM3NeT
and IceCube by considering the significance of a detection,
$\sigma\sim N_{sig}/\sqrt{N_{BG}}$.
Here, $N_{sig}$ is the number of signal events, while $N_{BG}$ is the
number of background events. The number of signal events in a time
interval $\Delta t$ scales as
$N_{sig}\propto A_{\nu}\cdot \Delta t\,,$
with $A_{\nu}$ as the intensity of the neutrino signal, which we
assume to be variable.
The number of background events simply proportional to
$N_{BG}\propto \Delta t\,,$
as the background does not change with time, apart from statistical
fluctuations. This results in a significance proportional to
$\sigma\propto A_{\nu}\cdot \sqrt{\Delta t}\,.$
If we compare the flaring state to the permanent state, we have
$A_{\nu}^{\rm flare}\approx x\cdot A_{\nu}^{\rm perm}$, 
with $A_{\nu}^{\rm flare}$ and $A_{\nu}^{\rm perm}$ as the
intensities of the flaring and permanent neutrino states, connected by
an intensity factor of $x\approx 10$.
The duration of the flare is $\Delta t^{\rm flare} \approx 3600\,{\rm s}\,,
$ and $\Delta
t^{\rm perm}$ is the observation time for a permanent flux. Now, we
can compare the significances of the flaring and the permanent state,
$\sigma_ {\rm flare}$ and $\sigma_{\rm perm}$:
\begin{eqnarray}
\frac{\sigma_{\rm flare}}{\sigma_{\rm perm}}&=&\frac{A_{\nu}^{\rm
      flare}}{A_{\nu}^{\rm perm}}\cdot
  \sqrt{\frac{\Delta{t}^{\rm flare}}{\Delta{t}^{\rm
        perm}}}= x\cdot  \sqrt{\frac{\Delta{t}^{\rm flare}}{\Delta{t}^{\rm
        perm}}}\\
&=&10^{-3}\cdot \left(\frac{x}{10}\right)\,
  \left(\frac{\Delta{t}^{\rm flare}}{3600\,{\rm s}}\right)^{1/2}\, \left(\frac{\Delta{t}^{\rm
        perm}}{1\,{\rm yr}}\right)^{-1/2}\,.\nonumber
\end{eqnarray}
This implies that the significance of a flaring state is a factor of
$10^{-3}$ lower compared to the observation of a permanent flux for
one year, considering that the flare's intensity is a factor of $10$
higher than the permanent emission. This leaves room for different
conclusions on whether a flaring state or the permanent neutrino
emission would be detected first:
\begin{itemize}
\item Consider that Pks2155-304 flares several times a year. If we
  assume a constant flare activity of $x=10$, the flaring state needs
  to be present more than $3\%$ of the time in one year - in that
  case, the flaring state can be observed before the permanent
  state. If the flaring rate is less, the permanent flux is likely to
  be observed earlier.
\item Single flares may be difficult to observed given the numbers
  above, unless there is no permanent emission, i.e.\ $\Delta t^{\rm
    flare} =\Delta t^{\rm perm}$. The optical depth we
  present is close to unity, and only a first-order approximation. If
  this optical depth drops far below unity most of the year, no
  permanent emission would be present and flaring states are likely to
  be observed first. In that case, we have $\sigma_{\rm flare}=x\cdot
  \sigma_{\rm perm}$, and selecting flaring states enhances the
    significance by a factor of $x$.
\end{itemize}

\section{Results and conclusions}
Very high-energy emission is observed from a growing number of
blazars. One possible emission scenario for the VHE component is
photon production in photo-hadronic interactions. In this paper, we investigate this
possibility by estimating the optical depth for proton-photon
interactions in the first strong shock of Pks2155-304. Using optical
photons, the optical depth in the low-state of Pks2155-304 is close to
unity, when assuming interactions with the optical photon field. The
observed correlation between optical and VHE wavelengths can therefore
be interpreted as a result of photo-hadronic interactions - an
increased optical flux leads to an enhancement of the proton-photon
optical depth and therefore to an increase of the VHE photon and
neutrino fluxes. At the same time, no correlation between VHE and
X-ray emission is necessary, since the optical depth for photohadronic
interactions at X-ray energies is too small to lead to significant
interactions. With IceCube extending the analysis of muon neutrinos to
the southern hemisphere, the prospects of detecting the flux from
Pks2155-304 are good within the first years of operation with
IceCube. While the sensitivity of Km3NeT will in general be
better, IceCube will be more sensitive to the important energy range,
e.g.\ $10^{8}$~GeV~$<E_{\nu}<10^{10}$~GeV and both instruments would
work complementary.

Flaring states provide the possibility of enhanced neutrino emission,
if the VHE radiation is due to $\pi^{0}-$decays. These flaring
states may in some cases be the only times of neutrino emission, if
the low-state provides too low optical depth for photo-hadronic
interactions. While Pks2155-304 provides reasonable optical depth for
neutrino production also in the low-state, this may not be the case
for other blazars. The stacking of neutrino data during observed
flaring states gives the unique opportunity to reduce the background
of atmospheric neutrinos in VHE neutrino telescopes like IceCube and
KM3NeT: Even a few neutrino events from such flares will be
significant. 
\subsection*{Acknowledgments}
We thank Francis Halzen, Alexander Kappes and Wolfgang Rhode for
inspiring discussions. Thanks to Berrie Giebels for providing the data
for the SED and lightcurve of Pks2155-304. Support for
the work of PLB has come from the AUGER membership and
 theory grant 05 CU 5PD 1/2 via DESY/BMBF and VIHKOS.

\end{document}